\documentclass[sigconf]{acmart}
\usepackage{dirtytalk}
\usepackage[english]{babel}

\AtBeginDocument{%
  }

\setcopyright{none}
\acmDOI{}
\acmISBN{}

\acmConference[LOCO '24]
{1st International Workshop on Low Carbon Computing (LOCO 2024)}
{December 3 2024}
{Glasgow/online}

\begin{document}

\title{Putting green software principles into practice}

\author{James Uther}
\email{james.uther@oliverwyman.com}
\affiliation{
	\institution{Oliver Wyman}
	\city{London}
	\country{UK}
}

\begin{abstract}
	The need and theoretical methods for measuring and reducing \coo\ emitted by computing systems are well understood, but real-world examples are still limited. We describe a journey towards green software for a live product	running on a public cloud. We discuss practical solutions found, in particular using the cost implications of serverless systems to drive efficiency. We end with some `green software' principles that worked well in this project.
\end{abstract}

\newcommand{\coo}{CO\textsubscript{2}}


\begin{teaserfigure}
	\includegraphics[width=\textwidth]{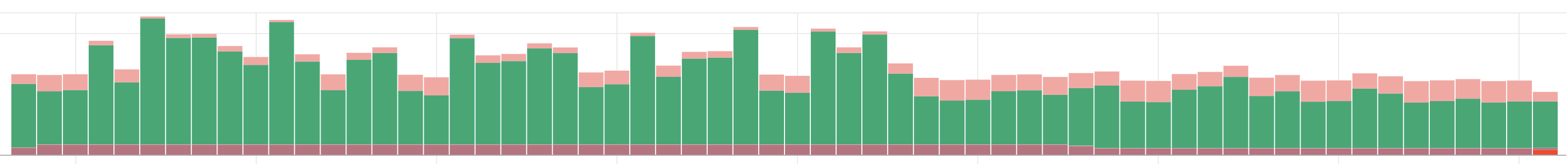}
	\caption{\coo\ footprint of a system over time.}
	\Description{Time series of \coo\ output.}
	\label{fig:teaser}
\end{teaserfigure}


\maketitle

\section{Introduction}

We recently built a mostly green-field product that involved substantial data processing, including developing and training various models and running production data pipelines using those models. Taking the advice \say{every job can be a climate job}~\cite{drawdownclimatejob} we sought from the outset to make the system as low-carbon as possible, given the business constraints.

We entered the project with a good understanding of green software principles, including the various sustainable architecture guidelines published by major hyperscalers \cite{AWSSusPillar,GCPSusPillar,MSSusPillar}, a "Green Software for Practitioners" \cite{GSFprac} badge, and personal motivation to make a difference where we were.

We started with some rough components that were running on a well known public cloud, and had reasonably free access to the cloud products as long as we coöperated with our corporate IT organisation. There was a gap between the proof of concept code and a system that could be run confidentially in production with orders of magnitude more data, so we were free to rewrite as necessary, within schedule constraints.

We now have a system and supporting infrastructure that could be considered `green'. In Section~\ref{sec:build} we will describe how we structured the project and systems, Section~\ref{sec:people} will discuss how the team found the process, and then we will offer some discovered principles in Section~\ref{sec:summary}. Note that finding these principles and building the system was an iterative process, with much learning and rebuilding.

\section{Building the Infrastructure\label{sec:build}}

We took a pragmatic definition of green software as \say{Does the job while producing minimal \coo e}. The `does the job' part is important, as it introduces other architecture drivers such as corporate policies, security, reliability, schedule, budget, and team capabilities. These are driven by business or institutional goals, and any green software principles or actions that make these harder for the team are likely to be ignored or resisted.

We could not identify any incentives to adopt green principles beyond personal ethics. While there are efforts to introduce legislation, policies and standards to drive the implementation of green software~\cite{secruk, eutaxonomy, secpolicy}, none applied to us at the time of writing, and in some cases are now unlikely to be finalised. This leaves green software in the odd space of being both desirable and at the same time all but valueless by the metrics of a company, and a distraction for most employees.

Thankfully, in the particular case of cloud computing the cost of the service can be used as a proxy for energy use, which can be related through grid intensity to an estimated \coo\ footprint. Cost is most definitely a good incentive for most businesses and thus employees. Low latency billing data from the cloud is watched carefully by cloud users, and proved invaluable in driving (and thus \coo ) reduction.

\subsection{The platform}

We needed to provide good \textit{infrastructure}~\cite{griffith2022electrify,saulinfra} that satisfies the user, while operating in a way that fulfils broader goals. We are used to compute infrastructure that allows us to get work done while transparently looks after security, IT policies, budget etc. We needed to build infrastructure that supported these corporate goals while producing as little \coo\ as possible.

Our strategy was to use products that are charged by usage and can automatically scale according to load (ideally to and from $0$). Such \textit{serverless} products are becoming ubiquitous as the cost and operational benefits can be considerable, particularly for spiky workloads. If the usage/billing data is available quickly they give an excellent feedback signal: Increased usage implies increased \coo\ and that usage is quickly reflected in the billing console. Actions to decrease cost involve reducing compute usage, which reduces \coo , linking good \coo\ management to good financial management.

Architectures for data intensive applications are well understood \cite{kleppmann2017designing} and have been packaged into various platforms, frameworks and libraries \cite{spark2010,ApacheBeam,moritz2018ray}. While we had business requirements regarding throughput and reliability, we also had to consider institutional knowledge and procurement policies. Luckily this left us with a few good options, and we chose a managed Spark \cite{spark2010} platform. The platform offered the ability to auto-scale clusters, and later a fully serverless cluster feature. We were also able to use efficient and cost effective ARM CPUs in most cases.

Serverless products at a large scale allow the provider optionality to more efficiently schedule workloads and increase hardware utilisation. At a small scale or with a known planned workload, this benefit vanishes, so relying on serverless options to reduce \coo\ (and cost) does not work well for on-prem compute.

\subsection{Software}

The platform guided users to use pyspark~\cite{pyspark}, which transforms python code into faster JVM operations~\cite{slowpython} that can be distributed on an auto-scaling cluster. Although Python was well known in the team, the need to transition from pandas~\cite{mckinney-proc-scipy-2010} to pyspark was more of a challange, but necessary as the production data was large enough to need distribution across a cluster. We did need to optimise the logic of the application itself, but could be guided by latency and cost considerations, also reducing \coo . While there is scope for further optimisation, the need to cater for diverse and unknown future maintainers limited our appetite for rewriting business logic in Rust.

We tried various ways of avoiding the \textit{thundering herd problem}~\cite{googlecron}. If everyone starts all batch jobs all at once (say 00:00 UTC, or on the hour), hyperscalers need to size capacity to those peaks leading to over-investment, higher embodied carbon, and the use of gas generators. Some cron implementations have wildcard values, allowing the data centre to schedule the job any time within given bounds. Mobile platforms overcome a related issue and save energy by providing a tolerance parameter within their timer APIs~\cite{appletimers}. Our platform provided only simple cron scheduling, so we are simply staring jobs at a fixed arbitrary time that is not on the hour. We hope this off-beat predictable schedule may help cloud providers manage hardware provisioning~\cite{adrianSun}. We would welcome being able to schedule jobs based on grid intensity (`when the wind blows'), but this is a feature best provided by the platform.

\subsection{Feedback}

We built dashboards to display daily costs associated with particular clusters and jobs, enabling us to highlight the need to further optimise parts of the system. It also helped prevent inefficient algorithms, architectures and libraries from gaining a foothold, as the costs associated quickly became apparent.

Section~\ref{sec:measure} describes our attempt to measure actual \coo , but it was not as effective to drive behaviour as cost. We also found our ability to measure increasingly limited as some workloads were moved to SAAS platforms and the compute hardware moved out of view.

\subsection{Power}
\begin{figure}[ht]
	\centering
	\includegraphics[width=\linewidth]{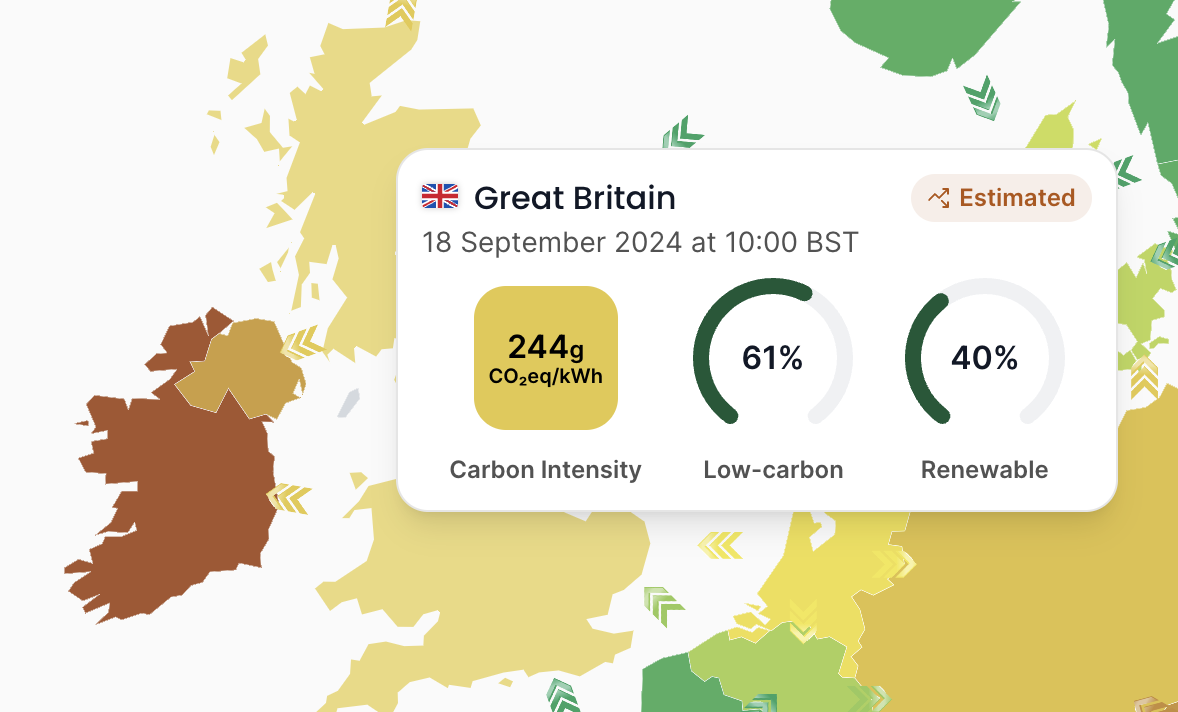}
	\caption{Live \coo\ intensity map\protect\footnotemark}
	\Description{A screenshot from a live \coo\ intensity map from \url{https://app.electricitymaps.com/map}.}
	\label{gridmap}
\end{figure}
\footnotetext[1]{\url{{https://app.electricitymaps.com/map}}}
We were able to use a cloud zone that claimed to be carbon neutral. We recognise that all hyperscalers make use of opaque carbon credit schemes~\cite{langer2023does,bjorn2022renewable} and a better assumption might be that the carbon intensity closely follows the regional grid intensity~\cite{electricitymaps} as shown in Figure~\ref{gridmap}.
Even this assumption is now doubtful as data centres are installing gas generators. Time-matched energy purchases (T-EACs)~\cite{teac,energytag} look like a way forward here and we look forward to seeing hyperscalers move in this direction.

\section{Measurement\label{sec:measure}}

Measurement is a foundation of good engineering. While we assumed the cloud bill was a reasonable proxy, we did try to get more detailed and true measure of our \coo\ footprint. There are now a few tools available to measure the \coo\ output of a system \cite{KeplerSlides,CloudCarbonFootprint,greenalg} and cloud providers also provide customers with carbon footprint reports. We have noticed that it's still unusual for SAAS platforms to do this, and estimating the footprint of their control planes and serverless workloads becomes impossible.

Our cloud platform provides a carbon usage dashboard, although the data trails by 3 months as electricity market intricacies are settled making it unhelpful for optimisation.
\begin{figure}[ht]
	\centering
	\includegraphics[width=\linewidth]{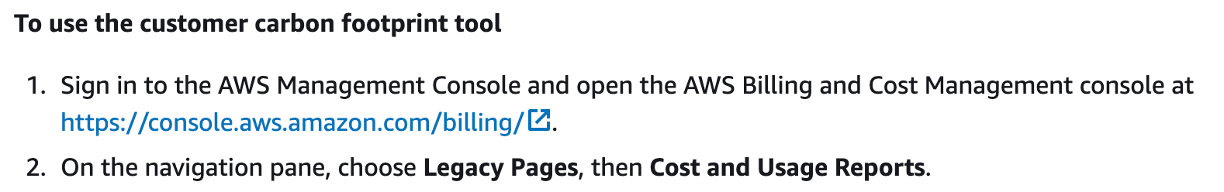}
	\caption{The \coo\ dashboard is considered legacy.}
	\Description{Instructions that say the dashboard we need is legacy.}
	\label{brokenawsco2instructions}
\end{figure}
The tool was also on a page of deprecated dashboards, had a minimun usage threshold to activate, and did not support reporting for sub-accounts, which was our situation. We view this lack of reliable, regular and \verb|actionable| information about our \coo footprint as a major gap in most serverless platform offerings, and encourage providers to provide an API for this purpose~\cite{cockroftqcon,keplerapi}.

As our workloads started to move to the managed serverless version of our spark platform we enquired whether that platform would support such an API. The provider admitted that \coo\ reporting was not on their roadmap.

\begin{figure}[ht]
	\centering
	\includegraphics[width=\linewidth]{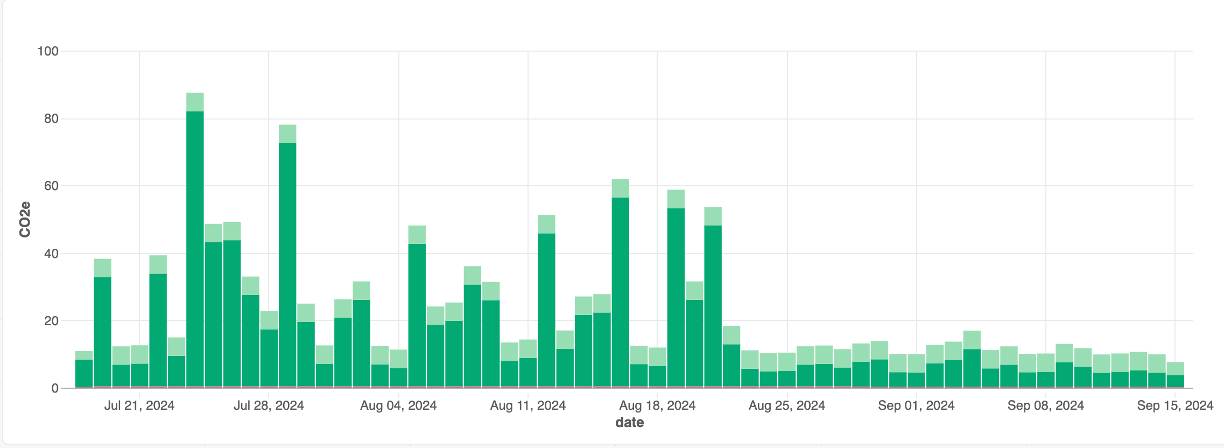}
	\caption{Our \coo\ footprint by day.}
	\Description{Instructions that say the dashboard we need is legacy.}
	\label{finalco2chart}
\end{figure}

We reverted to \verb|Cloud Carbon Footprint|~\cite{CloudCarbonFootprint} which has a methodology~\cite{ccfpMethodolgy} for converting billing data to estimated \coo\ footprint. Although usually used as an online system, we were abled to instead generate a lookup table\footnote{\url{https://www.cloudcarbonfootprint.org/docs/creating-a-lookup-table/}} that was imported into our platform, and used to augment the billing dashboard with our estimated footprint as in Figures~\ref{fig:teaser} and \ref{finalco2chart}.
The values here are based on estimates of the \coo\ footprint of the assumed hardware, given published figures and measurements of CPUs, memory, network etc, mixed with published PUE figures, and assumptions about management overhead. Kepler extends this with models trained on actual hardware~\cite{keplerMethodology}, but is focussed on kubernetes and thus not applicable here.

\section{People\label{sec:people}}

Having built foundations, we worked with others to build out the product. In a busy project there is little appetite for introducing risk and work for non-core goals. Thankfully the infrastructure of the project guided work to be low carbon while comfortably fulfilling the business needs. The data centre location (and grid intensity) was transparent. The billing data was vital for cost management. The platform scaled jobs to the lowest cost/\coo\ configuration automatically, or warned via billing when not being used effectively.

The project became a good learning opportunity for many. Often contributors were from a data science background which does not emphasise efficiency. Others were familiar with on-prem hardware where software efficiency has little impact on cost (which is carried in the initial purchase and fairly constant power bill). Some had not worked with data at this scale before. In all cases the move to a scalable serverless platform involved some learning around structuring solutions to enable distributed computing, and some lessons in efficiency and algorithm choice. It was not directly related to \coo\ footprint, but as the team learned the software improved and operating costs dropped. \coo\ dropped alongside.

\section{Summary\label{sec:summary}}

We would summarise the architecture principles and activities we used as:

\begin{itemize}
	\item \textbf{Infrastructure} which is intrinsically low carbon.
	\begin{itemize}
    \item \textbf{Renewable} electricity.
    \item \textbf{\#LightSwitchOps} - switch off unused resources.~\cite{zombies}
    \item Select the \textbf{right size} of resources.
  \end{itemize}
	\item \textbf{Feedback} by publishing usage (cost \& \coo ) data.
	\item \textbf{Optimise} the software.
\end{itemize}

These guided us to serverless solutions which was a new approach for some, but worth embracing for many reasons beyond their green credentials.

In our case, further reduction in \coo\ footprint is now largely dependant on our platform providers and their choice of electricity supply, cron services, scheduling algorithms, \coo\ footprint dashboards, supported software, hardware, cooling, etc. We decided that a large, dedicated organisation can build a more efficient platform than us, but we recognise it has left near-term progress in the hands of those organisations. In particular we would like to see better low latency reporting of our actual \coo\ impact.

\section{Citations and Bibliographies}

\bibliographystyle{ACM-Reference-Format}
\bibliography{ju_loco}

\end{document}